\documentclass[12pt]{article}
\usepackage{color}
\usepackage{graphicx}
\definecolor{violet}{rgb}{0.4,0,0.4}
\definecolor{vert}{rgb}{0,0.5,0.0}
\definecolor{navy}{rgb}{0.0,0.0,0.6}
\definecolor{orange}{rgb}{0.8,0.2,0.0}
\definecolor{bleu}{rgb}{0.3,0.0,0.8}

\def\vv{{\color{violet}v}} 

\def\phhi{{\color{violet}\varphi}}
\def\uu{{\color{violet}u}}

\def\sigm{{\color{red}\sigma}}

\def\calK{{\color{violet}{\cal K}}}

\def\Bl{{\color{black}{\ell }}}
\def\le{{\color{bleu}{e }}}

\def\ita{{\color{bleu} {\it a} }}

\def\alp{{\color{vert}\alpha}}
\def\bet{{\color{vert}\beta}}
\def\nn{{\color{vert}n}} 
\def\hnn{{\color{violet}\hat n}}
\def\bnn{{\color{violet}\overline n}}
\def\hbnn{{\color{violet}\hat {\overline n}}}

\def\kk{{\color{vert}k}}\def\ll{{\color{vert}l}}

\def\Elec{{\color{black}\rm E}}

\def\Te{{\color{orange}T}}
\def\pp{{\color{orange}p}}
\def\qq{{\color{vert}q}}
\def\aa{{\color{vert}a}}
\def\ee{{\color{orange}e}}
\def\mmu{{\color{orange}\mu}}

\def\mm{{\color{red}m}}\def\ee{{\color{red}e}}\def\jj{{\color{red}j}}
\def\hba{{\color{red}\hbar}}
\def\sigm{{\color{red}\sigma}}
  \def\VV{{\color{red}V}}
\def\DDelta{{\color{red}\Delta}}
\def\calH{{\color{red}{\cal H}}}
\def\dagh{{\color{violet}{\dagger}}}

\def\hHH{{\color{red}{\hat H}}}
\def\hpsi{{\color{violet}{\hat \psi}}}
\def\hcc{{\color{violet}{\hat c}}}
\def\gamh{{\color{violet}{\hat \gamma}}}
\def\hcalH{{\color{red}\hat{\cal H}}}

\def\calE{{\color{red}{\cal E}}}
\def\spose#1{\hbox to 0pt{#1\hss}}\def\lta{\mathrel{\spose{\lower 3pt\hbox
{$\mathchar"218$}}\raise 2.0pt\hbox{$\mathchar"13C$}}}  \def\gta{\mathrel
{\spose{\lower 3pt\hbox{$\mathchar"218$}}\raise 2.0pt\hbox{$\mathchar"13E$}}}

\def\Euro{\spose {\lower 2.5pt\hbox{${^{\bf =}}$}}{ C}}
\def\EE{{\color{red}\Euro}}

\def\bb{\large\color{black}  $ }  \def\fb{ $  }
\def\be{\large\begin{equation}\color{black} }
\def\fe{\end{equation}}
\def\rf{\color{black} (\ref }
\def\fr{)\,\color{navy} }

\def\spose#1{\hbox to 0pt{#1\hss}}\def\lta{\mathrel{\spose{\lower 3pt\hbox
{$\mathchar"218$}}\raise 2.0pt\hbox{$\mathchar"13C$}}}  \def\gta{\mathrel
{\spose{\lower 3pt\hbox{$\mathchar"218$}}\raise 2.0pt\hbox{$\mathchar"13E$}}}

\begin{document}

\title{\bf \color{navy}Effect of BCS pairing on entrainment in neutron
superfluid current in neutron star crust}

\author { {\bf
Brandon Carter$^{a}$,
Nicolas Chamel$^{a,b}$,
Pawel Haensel$^{a,b}$ }
\\ \hskip 1 cm\\   
$^{a}$ Observatoire de Paris, 92195 Meudon, France \\
$^{b}$ N. Copernicus Astronomical Center, Warsaw, Poland }

\date{\it 18 April, 2005}

\maketitle

\color{navy}

\bigskip
{\bf Abstract.}  The relative current density {\bb\nn^i\fb} of ``conduction''
neutrons in a neutron star crust beyond the neutron drip threshold can
be expected to be related to the corresponding particle momentum covector
{\bb\pp_i\fb} by a linear relation of the form {\bb\nn^i={\calK}^{ij}\pp_j\fb}
in terms of a physically well defined mobility tensor {\bb{\calK}^{ij}\fb}.
This result is describable as an ``entrainment'' whose effect -- wherever the
crust lattice is isotropic -- will simply be to change the ordinary neutron
mass {\bb\mm\fb} to a ``macroscopic'' effective mass {\bb\mm_\star\fb} such that in terms of
the relevant number density {\bb\nn\fb} of unconfined neutrons we shall have
{\bb{\calK}^{ij}=(\nn/\mm_\star)\gamma^{ij}\, .\fb} In a preceding work
based on a independent particle treatment beyond the Wigner-Seitz approximation,
using Bloch type boundary conditions
to obtain the distribution of energy
{\bb\calE_\kk\fb} and associated group velocity  {\bb \vv_{\kk}^i=\partial
\calE_\kk/\partial \hba \kk_i\fb} as a function of wave vector {\bb \kk_i\fb}, it
was shown that the mobility tensor would be proportional to a phase space volume
integral {\bb \calK^{ij}\propto\int {\rm d}^3\kk\,   \vv_{\kk}^i \vv_{\kk}^j\,\delta
\{\calE_\kk-\mmu\}\fb}, where {\bb \mmu\fb} is the Fermi energy. Using the
approach due to Bogoliubov, it is shown here that the effect of BCS pairing
with a superfluid energy gap {\bb \DDelta_{\rm F}\fb} and corresponding
quasiparticle energy function {\bb\EE_\kk=\sqrt{(\calE_\kk-\mmu)^2+
\DDelta_{\rm F}^2}\fb} will just be to replace the Dirac distributional integrand
by the smoother distribution in the formula {\bb \calK^{ij}\propto \int {\rm d}^3 \kk\,
\vv_{\kk}^i \vv_{\kk}^j\, \DDelta^{\,2}_{\rm F} /\EE^{\,3}_\kk\, .\fb} It is also shown
how the pairing condensation gives rise to superfluidity in the technical sense
of providing (meta) stability against resistive perturbations for a current that
is not too strong (its  momentum {\bb\pp_i\fb} must be small enough to give
{\bb 2|\pp_i\,\vv_{\kk}^i|<\EE^{\,2}_\kk/\vert \calE_\kk-\mmu\vert\fb} for all
modes). It is concluded that the prediction of a very large effective mass
enhancement in the middle layers of the star crust will not be significantly
effected by the pairing mechanism.

\vskip  1 cm

\section{Introduction}

A two fluid model of neutron star cores, of the kind that is commonly used in
hydrodynamical simulations has been developed in the past, assuming a neutron-proton-electron
composition and using a non relativistic treatment,
by the work of many authors \cite{ALS84,LM94}. In particular, as explained
by  Borumand {\it et al.} \cite{BJK96}, appropriate expressions have been obtained for
the relevant entrainment coefficients, relating the momentum of the neutron fluid to
the particle current of both neutrons and protons, in terms of the Landau parameters 
in the Fermi liquid theory. More recently, it has been shown by Comer {\it et al.}
 \cite{CJ2003,C2004} how the required entrainment coefficients, and the 
corresponding effective masses for the neutrons and the protons,  can be evaluated in 
a relativistic treatment -- such as is appropriate for deeper layers at densities 
substantially above the value (of about 10$^{14}$ g/cm$^3$) that characterises ordinary nuclear
matter -- using a relativistic $\sigma-\omega$ mean field model. The salient conclusion to be drawn
from all this work is that in these homogeneous fluid layers the effective mass
of the neutron will be significantly but not enormously reduced (by several tens of 
percent) below its ordinary bare mass value.

The present article is part of a newer program of work \cite{CCH,Ch2004} concerned 
with the previously unstudied problem of evaluating what turns out to be a much more
substantial effective mass modification that is obtained from the corresponding
entrainment coefficients in the crust at subnuclear densities, above the
neutron drip threshold (at about 10$^{11}$ g/cm$^3$). In these layers, relativistic corrections
are insignificant but the issue is complicated by the microscopic inhomogeneity
of the medium, in which some neutrons can still flow freely but protons are confined
to atomic nuclei. As in the outer neutron star crust where all nucleons
are bound, the nuclei of the inner crust will be liquid at high
temperatures but will be held by Coulomb forces in a crystalline solid lattice in the
relatively low temperature range (well below 10$^8$ degrees Kelvin) observed in
ordinary isolated neutron stars, which according to theoretical considerations  (see
e.g. \cite{Yakov2001}) should be attained within about a hundred years after the birth
of the star.  Such a low temperature regime will be maintained even in a
binary system involving accretion provided its rate does not exceed the typical order
of magnitude $\sim 10^{-10}$ $~{\rm M}_\odot$/yr  \cite{Miralda1990}. It has long
been generally recognised ~\cite{Sauls} that at such low temperatures (indeed all the
way up to 10$^9$ degrees Kelvin or more \cite{LombardoSch2001}) the neutrons will
form a BCS type condensate characterised by a superfluid energy gap, and -- for the
reasons discussed in the penultimate section of this article -- will therefore be 
able to flow through the lattice for a macroscopically long time \cite{PR95}, without 
resistive or viscous dissipation. The evaluation of the entrainment coefficients for 
such a superfluid flow is of particular astrophysical interest, because relative 
motion of the effectively free ``conduction'' neutrons through the inner crust lattice 
is believed to be an essential element in the mechanism responsible for observed 
pulsar frequency ``glitches''. 

In order to obtain the quantitative information that is needed, a first step has been 
the development by the present authors of a microscopic derivation \cite{CCH} of an 
appropriate two fluid model for the inner crust regime, on the basis of a simplified 
non-relativistic description of the underlying nuclear physics in which (as in the 
previously cited work \cite{BJK96,C2004} on the higher density regimes pertaining to 
deeper layers of the star) thermal corrections and the effects of superfluid pairing
were neglected. Recent numerical work \cite{Ch2004} has shown that this treatment 
implies an enormous enhancement by the entrainment of the effective mass of the free 
neutrons in the middle layers of the inner crust (a result that contrasts with the 
rather moderate diminution predicted for the effective neutron mass in the deeper 
core layers). The purpose of the present work is to evaluate the adjustments -- 
which turn out to be small -- that will result from allowance for the superfluid 
pairing of the neutrons.

Until now, quantum theoretical analysis of neutron superfluidity has
mainly concentrated on static configurations (in either infinite medium \cite{DHJ2004}
or inhomogeneous systems \cite{SGL, PBVB2001, MMM2004}), meaning states
for which no current is actually flowing relative to crust. Even at densities substantially
beyond the neutron drip threshold it should still be possible to obtain a reasonably
accurate description for the static case  by using the so called Wigner-Seitz approximation
that treats the neighbourhood of each ionic nucleus as if it were isolated in a sphere
whose diameter is determined by the nearest neighbour distance.
However for the treatment of more general -- stationary but
non static -- configurations involving relative conduction currents it is absolutely
necessary to use a more realistic description in which the artificial Wigner-Seitz
type boundary conditions are replaced by the natural Bloch type periodicity conditions
that would be desirable for higher accuracy even in the static case.

The use of appropriate Bloch type periodicity conditions is routine in terrestrial
 solid state physics \cite{Ashcroft}, but has so far been applied to neutron star
matter only in a simplified zero temperature independent particle treatment
\cite{CCH, Ch2004} (of
the kind applicable well below the Fermi temperature, but most appropriate for a
young neutrons star that has not yet fallen below the critical temperature for the
onset of superfluidity) for which the neutrons are considered to move as independent
particles without allowance for the pairing interactions responsible for the
superfluid energy gap that (in cool mature neutron stars) allows the currents to
persist.

A simplified treatment of this kind has been used to show that the middle layers of
a neutron star crust will be characterised by a very low value for the relevant
mobility tensor in the formula {\bb \nn^i=\calK^{ij}\pp_j\fb} for the current
{\bb \nn^i=\nn\, \bar \vv^i\fb} of unbound neutrons (which will be present above
the ``drip'' density of the order of 10$^{11}$ g/cm$^3$) with number density
{\bb \nn\fb}, mean velocity {\bb \bar \vv^i\fb} and momentum per neutron
{\bb\pp_i\fb}. Throughout this paper summation is understood over repeated covariant
and contravariant coordinate latin indices, for instance
{\bb \calK^{ij}\pp_j \equiv \sum_j \calK^{ij}\pp_j .\fb}
In the independent particle treatment, the mobility tensor was shown~\cite{CCH} to be given by a volume
integral over the space of Bloch momentum covectors {\bb \kk_i\fb} that is
expressible in terms of a Dirac distribution with support confined to the Fermi
surface -- where the relevant energy function {\bb\calE_{\kk\alp}\fb} with a band
index {\bb \alp, \fb} is equal to the chemical potential {\bb\mmu\fb} -- in the form
{\be \calK^{ij}={2\over (2\pi)^3}\sum_\alp \int \vv_{\kk\alp}^i \vv_{\kk\alp}^j\,
\delta\{\calE_{\kk\alp}-\mmu\}\,{\rm d}^3 k\, ,\label{00}\fe}
(in which, as throughout this work, we use braces -- as distinct from ordinary
brackets -- for functional dependence, in order to avoid possibly confusion with
simple multiplication) where the relevant group velocity distribution is given by
the usual formula
{\be \vv^i_{\kk\alp}=\frac{1}{\hba}{\partial\calE_{\kk\alp}\over\partial \kk_i}
\, .\label{32}\fe}

The main purpose of this article is to show how the preceding independent particle
treatment can be generalised to allow for BCS type pairing using an approach of the
kind pioneered by Bogoliubov. Since the relevant temperature range (substantially
below 10$^8$ degrees Kelvin) is very small compared with the critical value (of the
order of 10$^9$ degrees Kelvin) for the pairing condensation it will be justifiable
for us to continue to use the zero temperature limit, {\bb \Te=0\fb}, in which
thermal correction effects are entirely ignored. One of the main motivations for
this work is to check the robustness of the conclusions obtained from the simple
treatment described above, particularly the prediction of a very low value for the
mobility tensor, which is interpretable as meaning that the corresponding effective
mass {\bb\mm_\star=\nn/3\calK^i{_i}\fb} will become very large compared with the
ordinary neutron mass.

Our conclusion  is that as a first step towards a more accurate treatment, in cases
for which the superfluid pairing can be characterised just by a gap parameter
{\bb\DDelta_{\rm F}\fb}  the relevant integral over the Fermi surface  will  need
to be replaced by a phase space volume integral given in terms of the quasi-particle
energy
{\be \EE_{\kk\alp}=\sqrt{(\calE_{\kk\alp}-\mmu)^2+\DDelta_{\rm F}^2}\label{0a}\fe}
by the new formula
{\be \calK^{ij}={2\over( 2\pi)^3} \sum_\alp \int v_{\kk\alp}^i \vv_{\kk\alp}^j {\DDelta_{\rm F}^2
\over\EE_{\kk\alp}} \, {\rm d}^3\kk\, ,\label{0}\fe}
in which the expression for the group velocity {\bb \vv_{\kk\alp}^i\fb} is the same 
as in the absence of the pairing gap. It is the diminution of this group velocity that is
responsible for the enhancement of the effective mass, which on average should
therefore not be greatly affected by the phase space smearing effect produced by
the superfluid pairing. This comes from the fact that the expectation value in the superfluid phase
 of one particle quantities, namely the particle current density in the present work,
are not very different from their normal (non superfluid) value
as shown by Leggett \cite{Leggett75} in the context of superfluid helium 3.

Although the effect of the pairing is not so important for the evaluation of the
effective mass, it is of course important for the property of superfluidity itself.
Much of the contemporary literature on the underlying mechanisms for 
``superconductivity'' in astrophysical contexts deals only with purely static 
configurations, in which the essential question concerns the existence of a condensed 
state characterised by a finite energy gap. A secondary purpose of this article is to 
go back to the question of superconductivity in the strict technical sense, which 
refers not to static configurations but to stationary configurations involving the
relevant motion of a current of some kind -- electric in ordinary laboratory metals, 
but neutronic in the case of interest here. The essential issue is that of the (meta) 
stability of such a current against (small) perturbations of the kind that in the 
``normal'' case would produce resistive damping. The original defining property of a 
superconductor is that it should be able to support a current that will be 
characterised by such metastability provided it does not exceed some finite critical
value, beyond which ``normal'' dissipation will of course set in. It will be 
confirmed in the penultimate section of this article that within the framework of
 the simple theoretical model used for the present work this condition of 
superconductivity will indeed be satisfied, with a critical maximum current value 
that is estimated to be safely large compared with what is required for the relevant 
applications to astrophysical phenomena such as pulsar glitches.

\section{Hamiltonian for the independent particle limit}

The idea is to start on the basis of a second quantised formalism in terms of local
fermionic field annihilation and creation operators {\bb\hpsi\fb} and
{\bb\hpsi^\dagh\fb} depending on space position coordinates {\bb x^i\fb} in a unit
volume sample, and on a  spin variable {\bb\sigma\fb} taking values {\bb _\uparrow\fb}
and {\bb_\downarrow\fb} subject to  anticommutation rules of the usual form
{\be [\hpsi_{\sigma}\{ {\bf r} \}, \hpsi_{\sigma^\prime}\{ {\bf r^\prime} \}]_+=0
\, , \hskip 1 cm [\hpsi^\dagh_{\!\sigma}\{ {\bf r} \}, \hpsi^\dagh_{\!\sigma^\prime}
\{ {\bf r^\prime} \}]_+ =0\, ,\label{1} \fe}
{\be [\hpsi^\dagh_{\!\sigma}\{ {\bf r} \}, \hpsi_{\sigma^\prime}\{ {\bf r^\prime}\}
]_+=\delta_{\sigma\sigma^\prime}\delta\{ {\bf r} , {\bf r^\prime}\}\, ,\label{2}\fe}
using a quadratic Hamiltonian operator of the form
{\be \hHH=\hHH_{_{\rm ind}}+\hHH_{_{\rm int }}\, ,\hskip 1 cm \hHH_{_{\rm ind}}=
\hHH_{_{\rm kin}}+\hHH_{_{\rm pot}}\, ,\label{3}\fe}
in which the interaction term {\bb\hHH_{_{\rm int}}\fb} will be absent in the
independent particle limit corresponding to the kind of model used~\cite{CCH} in our
preceding first quantised treatment.

In this independent particle limit, only the kinetic
and potential contributions are present and will be assumed to be given (neglecting possible
spin dependence for simplicity) by integrals over the unit volume sample under
consideration of the form
{\be \hHH_{_{\rm kin}}=\int{\rm d}^3 r \, \hcalH_{_{\rm kin}}\{ {\bf r} \}\, ,\hskip
1 cm \hcalH_{_{\rm kin}}\{ {\bf r} \}=\sum_\sigma\hpsi^\dagh_{\!\sigma}\{{\bf r}\}
\calH_{_{\rm kin}}\hpsi_\sigma\{{\bf r} \}\, ,\label{7}\fe}
{\be \hHH_{_{\rm pot}}=\int{\rm d}^3 r \, \hcalH_{_{\rm pot}}\{ {\bf r} \}\, ,
\hskip 1 cm \hcalH_{_{\rm pot}}\{ {\bf r} \}=\VV\{ {\bf r} \}  \sum_{\sigma}
\hpsi^\dagh_{\!\sigma}\{ {\bf r} \} \hpsi_{\sigma}\{ {\bf r} \}\, ,\label{4}\fe}
where {\bb\calH_{_{\rm kin}}\fb} is a self adjoint differential
operator in the category specified in terms of a gauge covector
{\bb a_i\fb} by an expression of the familiar form
{\be \calH_{\aa}=-\gamma^{ij}(\nabla_i+i\aa_i){1\over 2 \mm^{_\oplus}\{  {\bf r} \}}
(\nabla_i+i\aa_i)\, \label{8}\fe}
in which {\bb \gamma^{ij}\fb} is the Euclidean space metric and
{\bb\mm^{_\oplus}\{ {\bf r}\}\fb} is interpretable as a microcopic effective mass, which
is usually found to have smaller values inside crustal nuclei \cite{SGL}.
The covector with components {\bb \aa_i\fb} is
a gauge field allowing for the possibility of adjustment of the phases
of the field operators {\bb \hpsi_{\!\sigma}\{{\bf r}\}\fb}. In
applications to particles with non zero electric charge ( {\bb \ee\fb} say)
such as the electrons in an ordinary terrestrial superconductor or the
protons in the deeper layers of a neutron star, the presence of such a field
(taking the form {\bb \aa_i=\ee A_i\fb}) would be necessary for the treatment
of magnetic effects, but in the uncharged case of the crust neutrons with
which we are concerned here it will always be possible to work in the
standard gauge for which this covector is simply set to zero,
{\bb \aa_i=0\fb},  which means that we simply take
{\be\calH_{_{\rm kin}}=\calH_{0}\, .\label{9}\fe}

The potential {\bb \VV\{ {\bf r} \} \fb} and
the microscopic effective mass {\bb \mm^{_\oplus}\{ {\bf r} \}\,  \fb}
(as those deduced from contact two body interactions of the Skyrme type in the
Hartree Fock approximation)
are supposed to represent the averaged effect
on the neutrons of the nuclear medium and in particular of the ionic lattice.
A periodic crystalline type lattice will be assumed, which
implies that these functions should be invariant with respect to any lattice
translation vectors :
{\be\VV\{ {\bf r}+\Bl^\ita {\bf \le}_\ita \} =\VV\{ {\bf r} \}\, , \hskip 1cm
 \mm^{_\oplus}\{ {\bf r}+\Bl^\ita {\bf \le}_\ita \}   = \mm^{_\oplus}\{ {\bf r} \}\label{6}\fe}
for any triad of integers {\bb \Bl^\ita \fb}  ({\bb \ita=1,2,3\fb})
in which the lattice basis vectors {\bb {\bf \le}_\ita\fb}
may be interpreted as representing the interionic spacing in the solid case that will
be relevant at very low temperature, but should in principle be taken to be much
larger (so as to generate a  giant cell interpretable as a typical mesoscopic average
over a locally disordered configuration) for applications above the relevant melting
temperature, at which it is to be expected that (unlike the weaker electron pairing
mechanism in ordinary terrestrial superconductors) the superfluid neutron pairing
mechanism will still be intact.

The mass function {\bb\mm^{_\oplus}\{ {\bf r} \}\fb} in the specification of the
kinetic contribution to the independent Hamiltonian will also be involved in
the specification of the corresponding neutron current density operators, which will
 be given, for each value of the spin variable {\bb\sigma\fb}, by
{\be \hnn^{\, i}_\sigma\{{\bf r} \}={\hba\over 2{\rm i} \mm^{_\oplus} \{ {\bf r} \}}
\gamma^{ij}\left(\hpsi^\dagh_{\!\sigma}\{{\bf r}\}\nabla_{\!j}
\hpsi_{\sigma}\{{\bf r}\}-(\nabla_{\!j}\hpsi^\dagh_{\!\sigma})
\hpsi_{\sigma}\{{\bf r}\}\right)\, .\label{10}\fe}
One of the main objectives of the present work is to obtain a practical way of
evaluating the mean value of the total current, as given by the
space averaged operator
{\be  \hbnn^{\, i}= \sum_\sigma\hbnn^{\, i}_\sigma\, , \hskip 1 cm
 \hbnn^{\, i}_\sigma=\int{\rm d}^3 r\, \hnn^{\, i}_\sigma\, ,\label{26}\fe}
as a function of the associated momentum in a stationary state that is non static
(and therefore non isotropic, since the mean current will characterise a preferred
direction) but uniform of a mesoscopic volume, meaning one that is large compared
with the interionic spacing but small compared with the macroscopic lengthscales
characterising the star crust thickness or even the intervortex separation.

It is to be noted for future reference that this current can be used to express
the adjustment that will be required in cases when it turns out to be more convenient
to work with the gauge adjusted operator {\bb \calH_{\aa}\fb} rather than
{\bb \calH_{0}\fb} in the kinetic contribution \rf{9}\fr: it can be seen that this
kinetic contribution will be given in the small {\bb\aa\fb} limit by
{\be \hHH_{_{\rm kin}}+\aa_i \, \hbnn^{\, i}
=\sum_\sigma\int{\rm d}^3 r \, \hpsi^\dagh_{\!\sigma}\{{\bf r}\}
\calH_{\aa}\hpsi_\sigma\{{\bf r} \}
+{\cal O}\{|\aa|^2\}\, ,\label {26b}\fe}
subject to the usual assumption that we are using periodic boundary conditions
to get rid of a boundary term produced by an integration by parts using Green's
theorem.

\section{Representation by Bloch states}

Subject to the usual Bloch type boundary conditions for a mesoscopic material
sample of parallelopiped form  -- with a unit volume that is taken to be very large
compared with the elementary lattice cells under consideration -- the independent
particle Hamiltonian will determine a complete orthonormal set of single particle states
{\bb \phhi_{\kk\alp}\{{\bf r}\}\fb}, labelled by a wave covector {\bb \kk_i\fb}
taking discrete values on a fine mesh inside the first Brillouin zone and a band
index {\bb \alp \fb}, satisfying the Floquet-Bloch theorem \cite{Ashcroft}
{\be \phhi_{\kk\alp}\{{\bf r}\}=\uu_{\kk\alp}\{{\bf r}\}\, {\rm e}^{{\rm i}\, \kk
\cdot \bf r} \, ,\label{12}\fe}
using the abbreviation {\bb \kk\cdot {\bf r}=\kk_i \, x^i\fb}, where
{\bb\uu_{\kk\alp}\{{\bf r}\}\fb} satisfies the ordinary lattice periodicity
conditions
{\be\uu_{\kk\alp}\{{\bf r}+\Bl^\ita {\bf \le}_\ita \}=\uu_{\kk\alp}\{ {\bf r} \}
\, .\label{13}\fe}
These wave functions are normalised as follows (using {\bb ^\ast\fb} to indicate
complex conjugation):
{\be \int{\rm d}^3 r\,
\phhi^\ast_{\kk\alp}\{{\bf r}\}\phhi_{\ll\bet}\{{\bf r}\}
=\delta_{\kk\ll}\delta_{\alp\bet}\, .\label{14}\fe}
Setting the Bloch wave vector {\bb \kk_i\fb} in place of {\bb \aa_i\fb} in the
definition \rf{8}\fr the eigenvalue equation can be usefully rewritten in terms of
the ordinarily periodic functions {\bb \uu_{\kk\alp} \fb} as
{\be \big(\calH_\kk+\VV\big)\uu_{\kk\alp}\{{\bf r}\}=\calE_{\kk\alp}
\uu_{\kk\alp}\{{\bf r}\}\, .\label{16}\fe}
From the spin independence of the potential (by neglecting spin-orbit
coupling terms), the phases
can be chosen in such a way that we shall have
{\be \phhi^\ast_{\kk\alp}\{{\bf r}\}=\phhi_{-\kk \alp}\{{\bf r}\}\, ,\hskip 1 cm
 \uu^\ast_{\kk\alp}\{{\bf r}\}=\uu_{-\kk\alp}\{{\bf r}\}\, .\label{17} \fe}

These Bloch states may be employed in the usual way as a basis for the
specification of corresponding  position independent annihilation and creation
operators, {\bb \hcc_{\sigma\alp\kk} \fb} and  {\bb \hcc^\dagh_{\!\sigma\kk\alp}\fb},
subject to anticommutation relations of the standard form
{\be [\hcc_{\sigma\kk\alp}, \hcc_{\sigma^\prime\ll\bet}]_+ =0\, , \hskip 1 cm
[\hcc^\dagh_{\!\sigma\kk\alp}, \hcc^\dagh_{\!\sigma^\prime\ll\bet}]_+
=0\, ,\label{19} \fe}
{\be [\hcc^\dagh_{\!\sigma\kk\alp}, \hcc_{\sigma^\prime\ll\bet}
]_+=\delta_{\sigma\sigma^\prime}\delta_ {\kk\ll}\delta_{\alp\bet}\, ,\label{20}\fe}
in terms of which the original position dependent annihilation and creation operators
will be given by
{\be \hpsi_\sigma\{{\bf r}\}=\sum_{\kk,\alp} \phhi_{\kk\alp}\{{\bf r}\}
\hcc_{\sigma\kk\alp} \, ,\hskip 1 cm
\hpsi^\dagh_\sigma\{{\bf r} \}=\sum_{\kk, \alp} \phhi^\ast_{\kk\alp}\{{\bf r}\}
\hcc^\dagh_{\!\sigma\kk\alp}\, .\label{21}\fe}

It is evident just from the orthonormality conditions \rf{14}\fr that the spin
dependent number density operator defined by
{\be \hnn_\sigma\{ {\bf r}\}=\hpsi^\dagh_{\!\sigma}
\{{\bf r}\}\hpsi_{\sigma}\{{\bf r}\}\, . \label{5}\fe}
will have a space integral
{\be \hbnn_\sigma=\int{\rm d}^3 r\, \hnn_\sigma\{{\bf r}\}\, ,\label{22}\fe}
(over the unit volume sample under consideration) that will be given by
{\be \hbnn_\sigma
=\sum_{\kk, \alp} \hnn_ {\sigma\kk\alp}\, ,\hskip 1 cm \hnn_ {\sigma\kk\alp}
=\hcc^\dagh_{\!\sigma\kk\alp}\hcc_{\sigma\kk\alp}
\, .\label{30b}\fe}
It can similarly be seen from the defining conditions \rf{21}\fr that integrated
value of the independent particle contribution,
{\be \hHH_{_{\rm ind}}=\int{\rm d}^3 r\, \hcalH_{_{\rm ind}}\{{\bf r} \}\, ,
\hskip 1 cm \hcalH_{_{\rm ind}}\{ {\bf r} \}=\hcalH_{_{\rm kin}}\{{\bf r} \}+
\hcalH_{_{\rm pot}}\{{\bf r} \}\, ,\label{22a}\fe}
which is interpretable in the absence of the interaction contribution as the total
energy operator, will be expressible in standard form as
{\be  \hHH_{_{\rm ind}}=\sum_{\sigma, \kk, \alp}
\calE_{ \alp \kk}\,\hnn_ {\sigma\kk\alp}\, ,\label{23}\fe}
where {\bb \hnn_ {\sigma\kk\alp}\fb} is the Bloch wave vector dependent particle number
density operator given by \rf{30b}\fr.

To get an analogous formula for the mean current (over the unit volume sample
under consideration) as given by the operator \rf{26}\fr, we
take its expectation value
{\be \langle|\hbnn^i_\sigma|\rangle=\int{\rm d}^3 r\, \langle |\hnn^i_\sigma
\{{\bf r}\}|\rangle\, ,\label{27}\fe}
for a state
{\bb |\,\rangle\fb} satisfying the simplicity condition that except for the
diagonal contributions characterised by {\bb \sigma^\prime=\sigma, \,\ll_i
=\kk_i \fb} and {\bb \alp=\bet\fb} the contributions of the expectation values
{\bb \langle|\hcc^\dagh_{\!\sigma\kk\alp}\hcc_{\sigma^\prime\ll\bet} |\rangle\fb}
will vanish -- or  be negligible to the order of approximation under consideration --
it can be seen that we shall obtain the formula
{\be \langle |\hbnn^i_\sigma|\rangle=\sum_{\kk, \alp} \langle| \hnn_{\sigma\kk\alp}
|\rangle \vv^i_{\kk\alp}\, ,\label{27a}\fe}
in which the relevant velocity will be given by
{\be \vv^i_{\kk\alp}=\int{\rm d}^3 r\, {\hba\over 2{\rm i}\mm^{_\oplus}\{{\bf r} \}}
\gamma^{ij}\Big(\phhi^\ast_{\kk\alp}\{{\bf r}\}\nabla_{\!j}\phhi_{\kk\alp}\{{\bf r}
\}-\phhi_{\kk\alp}\{{\bf r} \}\nabla_{\!j}\phhi^\ast_{\kk\alp}\{{\bf r} \}\Big)
\, .\label{27b}\fe}
This  expression \rf{27b}\fr for the velocity vector
{\bb \vv^i_{\kk\alp}\fb} can easily be shown to be mathematically equivalent to the well
known, albeit less intuitively obvious, group velocity formula that is given in terms of
the single particle energy introduced in \rf{16}\fr by \rf{32}\fr.

\bigskip
\section{Characterisation of  conducting reference state}
\medskip

The (zero temperature) states in which we are interested are those that minimise the
expected total energy {\bb \langle|\hHH|\rangle\fb} subject not only to the usual
constraint that there should be a fixed given value of the corresponding total expected
particle number
{\be \langle |\hbnn|\rangle=
\sum_\sigma \langle |\hbnn_\sigma|\rangle\, ,\label{28}\fe}
but also, since we are concerned with non-static -- conducting -- stationary
configurations, to the requirement that there should also be a fixed given value of
the expected total
{\be \langle |\hbnn^{\, i}|\rangle=
\sum_\sigma \langle |\hbnn^{\ i}_\sigma|\rangle\, ,\label{29}\fe}
of the current defined by \rf{26}\fr.

Imposing these constraints by the introduction of corresponding Lagrange multipliers 
{\bb \mmu\fb} and {\bb \pp_i \fb}, the problem will effectively be that of 
unconstrained  minimisation of the combination
{\be \langle|\hHH_{\{\pp\}}^\prime|\rangle=\langle|\hHH|\rangle
-\mmu\langle |\hbnn|\rangle-\pp_i
\langle |\hbnn^{\, i}|\rangle\, ,\label{33}\fe}
in which we introduce the notation
{\be \hHH_{\{\pp\}}^\prime= \hHH^\prime-\pp_i\, \hbnn^{\, i}\, ,\hskip 1 cm
 \hHH^\prime=\hHH-\mmu \hbnn\, .\label{33b}\fe}
In the absence of the pair coupling term {\bb\hHH_{_{\rm int}}\fb}, the quantity to be
minimised reduces to the form {\bb\langle|\hHH_{_{\rm ind}\{\pp\}}^\prime|\rangle\fb}
with
{\be \hHH_{_{\rm ind}\{\pp\}}^\prime= \hHH^\prime_{_{\rm ind}}
-\pp_i\, \hbnn^{\, i}\, ,\hskip 1 cm
 \hHH^\prime_{_{\rm ind}}=\hHH_{_{\rm ind}}-\mmu \hbnn\, .\label{33c}\fe}
It can be seen from \rf{9}\fr and \rf{26b}\fr that, for a small value,
{\be \pp_i = \hba \qq_i\, , \fe}
of the momentum,  its effect will be given to first order in the magnitude {\bb|\qq|\fb}
of the corresponding wave number covector just by substituting the gauge adjusted
operator {\bb \calH_{-\qq}\fb} in place of  {\bb \calH_{0}\fb} in the relevant
differential formulae. Thus,  in particular, it can be seen that the appropriate
modification of the eigenvalue equation \rf{16}\fr for the required replacements
{\bb\calE_{\{\pp\}\kk\alp}\fb} and {\bb\uu_{\{\pp\}\kk\alp}\fb}  of
{\bb\calE_{\kk\alp}\fb} and {\bb\uu_{\kk\alp}\fb} will be just the substitution
of {\bb \calH_{\kk-\qq}\fb} for  {\bb \calH_{\kk}\fb}, which evidently means that to
this first order of accuracy we shall have
{\be \calE_{\{\pp\}\kk\alp}=\calE_{(\kk-\qq)\alp}\, ,\fe}
and
{\be \uu_{\{\pp\}\kk\alp}=\uu_{(\kk-\qq)\alp}\, .\fe}
One thereby obtains the  formula
{\be \phhi_{\{\pp\}\kk\alp}\{{\bf r}\}=\, {\rm e}^{{\rm i}\, \qq
\cdot \bf r}\phhi_{(\kk-\qq)\alp}\{{\bf r}\}\, ,\fe}
for the corresponding modification of the single particle states \rf{12}\fr,
which in turn, by the analogue of \rf{21}\fr, determine correspondingly adjusted
annihilation and creation operators,  {\bb \hcc_{\{\pp\}\sigma\kk\alp} \fb} and
{\bb \hcc^\dagh_{\!\{\pp\}\sigma\kk\alp}\fb}, in terms of which \rf{30b}\fr can be
rewritten in the equivalent form
{\be \hbnn_\sigma
=\sum_{\kk, \alp} \hnn_ {\{\pp\}\sigma\kk\alp}\, ,\hskip 1 cm \hnn_{\{\pp\}\sigma\kk\alp}
=\hcc^\dagh_{\!\{\pp\}\sigma\kk\alp}\hcc_{\{\pp\}\sigma\kk\alp}
\, .\label{30e}\fe}
It can thus be seen that, as the analogue of \rf{23}\fr, the effective Hamiltonian
{\bb \hHH_{\{\pp\}_{\rm ind}}^\prime\fb} will be given by the formula
{\be \hHH_{\{\pp\}_{\rm ind}}^\prime=\sum_{\sigma,\kk,\alp}
\calE^\prime_{\{\pp\}\kk\alp}\,\hnn_ {\{\pp\}\kk\alp}
\, ,\label{34}\fe}
with
{\be \calE^\prime_{\{\pp\}\kk\alp}=\calE^\prime_{(\kk-\qq)\alp}\, ,\label{35}\fe}
to the linear order accuracy in {\bb \pp\fb} with which we are working. At this order,
it can be seen from  \rf{32}\fr that \rf{35}\fr may be rewritten as
{\be \calE^\prime_{\{\pp\}\kk\alp}=\calE^\prime_{\kk\alp}-\pp_i\, v_{\kk\alp}^i
\, ,\hskip 1 cm \calE^\prime_{\kk\alp}=\calE_{\kk\alp}-\mmu\, .\label{35a}\fe}

The expectation value of the quantity given by \rf{34}\fr will evidently be minimised
by a reference state vector {\bb |\,\rangle= |_{\{\mmu,\pp\}}\rangle  \fb} that is
chosen (as a function of the multipliers {\bb \mmu\fb} and {\bb\pp_i\fb})
in such a way that the expectation
{\bb \langle_{\{\mmu,\pp\}}|\hnn_{\{\pp\}\sigma\kk\alp}|_{\{\mmu,\pp\}}\rangle\fb}
has its maximum value, namely 1, whenever {\bb\calE^\prime_{\{\pp\}\kk\alp}\fb} is negative,
and its minimum value, namely zero, whenever {\bb\calE^\prime_{\{\pp\}\kk\alp}\fb} is
positive. It can thus be seen from \rf{35}\fr that the effect of the current will
consist just of a uniform shift of the distribution in momentum space by an amount
given by the infinitesimal momentum covector {\bb \pp_i\fb}. Such a state is
characterised by the conditions
{\be \hnn_{\{\pp\}\sigma\kk\alp}|_{\{\mmu,\pp\}}\rangle
=\nn_{\{\pp\}\sigma\kk\alp}|_{\{\mmu,\pp\}}\rangle\, .\label{36a}\fe}
with the eigenvalues given as a Heaviside distribution by
{\be \nn_{\{\pp\}\sigma\kk\alp}=\vartheta\{-\calE^\prime_{\{\pp\}\kk\alp}\}
\, .\label{36b}\fe}

It can be seen that this state  {\bb |_{\{\mmu,\pp\}}\rangle  \fb} satisfies the
condition for applicability of the analogue of \rf{27a}\fr, and hence that the
expected mean current value
{\be \bnn^i_\sigma=\langle_{\{\mmu,\pp\}}|\hbnn^i_\sigma|_{\{\mmu,\pp\}}\rangle
\label{36h}\fe}
will be given  (in accordance with our previous evaluation~\cite{CCH} in a first
quantised framework) by
{\be \bnn^i_\sigma=\sum_{\kk,\alp} v^i_{\kk\alp}\,\vartheta
\{-\calE^\prime_{\{\pp\}\kk\alp}\}\, .\label{36g}\fe}
In the linearised weak current limit with which we are working, it can be seen from
\rf{32}\fr and \rf{35a}\fr that this will be expressible to first order in terms of the
static limit value,
{\be \nn_{\sigma\kk\alp}=\vartheta\{\mmu-\calE_{\kk\alp}\}\, ,\label{36c}\fe}
of the distribution \rf{36b}\fr as
{\be  \bnn^i_\sigma=\pp_j\sum_{\kk,\alp}\frac{ \nn_{\sigma\kk\alp}    }{\hba^2}
{\partial^2\calE_{\kk\alp}\over\partial \kk_i\partial \kk_j} \, .\label{36d}\fe}

\section{Bogoliubov treatment of pairing}

Up to this point what has been done in  the present article is just to translate the 
work of our preceding article~\cite{CCH} from first quantised to second quantised 
formalism. The motivation for this translation is that a second quantised treatment 
is indispensible for the next step, which is to go beyond the independent particle 
model used in the preceding work by including allowance for pairing interactions.

In the inner crust in which we are concerned here, dripped neutrons are expected to 
be paired in spin singlet states \cite{Sauls}, as the usual Cooper pairs of electrons 
in terrestrial superconductors. A standard way of allowing for pairing interaction in 
a mean field model (assuming for simplicity a contact two body interaction as it is 
the case for conventional superconductor \cite{Ketterson} and a common practice in 
nuclear physics \cite{BH2003})  is thus to take the interaction contribution in 
\rf{3}\fr to have the form
{\be \hcalH_{_{\rm int}}\{{\bf r}\}=\DDelta\{{\bf r}\}
\hpsi^\dagh_{\!\uparrow}\{{\bf r}\}\hpsi^\dagh_{\!\downarrow}\{{\bf r}\}
+\DDelta^\ast\{{\bf r}\}\hpsi_\downarrow\{{\bf r}\}\hpsi_\uparrow\{{\bf r}\}
\, ,\label{37}\fe}
where {\bb\DDelta\{{\bf r}\}\fb} is a position dependent complex potential that, in 
a ``self consistent'' model, should be expressible in terms of the
abnormal density expectation value
{\bb \langle|\hpsi_\downarrow\{{\bf r}\}\hpsi_\uparrow\{{\bf r}\}|\rangle\fb} in the relevant
reference state {\bb |\rangle\, .\fb}

The mean complex phase of the function {\bb\DDelta\{{\bf r}\}\fb} is subject to an 
indeterminacy that can be resolved by fixing the phase in the specification of the 
wave operators. In a static configuration one would expect that this coupling 
potential {\bb\DDelta\{{\bf r}\}\fb} would share the ordinary lattice periodicity property
\rf{6}\fr 
and moreover that the phase should be adjustable in such a way as to ensure that
{\bb\DDelta\fb} becomes
real.

Instead of using the representation \rf{12}\fr in terms of the simple Bloch wave 
functions {\bb\phhi_{\!\kk\alp}\{{\bf r}\}\fb}, in the approach introduced by 
Bogoliubov one seeks a more general representation whereby the single component Bloch 
waves are replaced by two component Bloch functions with components
{\bb\phhi^{_0}_{\kk\alp}\{{\bf r}\}\fb} and  {\bb\phhi^{_1}_{\kk\alp}\{{\bf r}\}\fb} 
that are characterised with respect to corresponding ordinarily periodic functions 
{\bb\uu^{_0}_{\kk\alp}\{{\bf r}\}\fb}, and {\bb\uu^{_1}_{\kk\alp}\{{\bf r}\}\fb} by
{\be \left({\phhi^{_0}_{\kk\alp}\{{\bf r}\}\atop \phhi^{_1}_{\kk\alp}\{{\bf r}\}}
\right)={\rm e}^{{\rm i}\, \kk\cdot\bf r}
\left({\uu^{_0}_{\kk\alp}\{{\bf r}\}\atop \uu^{_1}_{\kk\alp}\{{\bf r}\}}
\right)\, .\label{46}\fe}
These functions are used for replacing the original representation \rf{12}\fr by a 
mixed particle-hole representation involving a new set of position independent 
quasi-particle annihilation and creation operators
{\bb \gamh_{\sigma\kk}\fb} and {\bb\gamh^\dagh_{\!\sigma\kk}\fb}
in terms of which we shall have
{\be  \hpsi_{\!\uparrow}\{{\bf r}\}=\sum_{\kk,\alp}\left(\phhi^{_0}_{\kk\alp}
\{{\bf r}\}\gamh_{\uparrow\kk\alp}-\phhi^{_1\ast}_{\!\kk\alp}\{{\bf r}\}
\gamh_\downarrow{^{\!\dagh}_{\kk\alp}}\right)\, ,\label{47}\fe}
and
{\be  \hpsi_{\!\downarrow}\{{\bf r}\}=\sum_\kk\left(\phhi^{_0}_{\kk\alp}
\{{\bf r}\}\gamh_{\downarrow\kk\alp}+\phhi^{_1\ast}_{\!\kk\alp}\{{\bf r}\}
\gamh_\uparrow{^{\!\dagh}_{\kk\alp}}\right)\, ,\label{48}\fe} where the new operators
satisfy anticommutation relations of the standard form
{\be [\gamh_{\sigma\kk\alp}, \gamh_{\sigma^\prime\ll\bet}]_+
=0\, , \hskip 1 cm
[\gamh^\dagh_{\!\sigma\kk\alp}, \gamh^\dagh_{\!\sigma^\prime \ll\bet}]_+
=0\, ,\label{60} \fe}
{\be [\gamh^\dagh_{\!\sigma\kk\alp}, \gamh_{\sigma^\prime\ll\bet}
]_+=\delta_{\sigma\sigma^\prime}\delta_{\kk\ll}\delta_{\alp\bet}\, .\label{61}\fe} 
As a result, consistency with \rf{1}\fr and \rf{2}\fr entails the relations:
 {\be [ \hpsi^\dagh_{\!\uparrow}\{{\bf r}\}, \hpsi_{\!\uparrow}\{{\bf r}^\prime\}]_+
=\delta\{{\bf r},{\bf r}^\prime\} = \sum_\kk \phhi^{_0\ast}_{\kk\alp}\{{\bf r}\} 
\phhi^{_0}_{\kk\bet}\{{\bf r}^\prime\} +\phhi^{_1\ast}_{\kk\alp}\{{\bf r}^\prime\}
 \phhi^{_1}_{\kk\alp}\{{\bf r}\}\, ,\fe}
{\be [\hpsi^\dagh_{\!\downarrow}\{{\bf r}\}, \hpsi^\dagh_{\!\uparrow}\{
{\bf r}^\prime\}]_+ =0= \sum_\kk \phhi^{_1}_{\kk\alp}\{{\bf r}\} 
\phhi^{_0\ast}_{\kk\alp}\{ {\bf r}^\prime \} - \phhi^{_0\ast}_{\kk\alp}\{{\bf r}\} 
\phhi^{_1}_{\kk\alp}\{ {\bf r}^\prime \} \, . \fe}

The purpose of such a Bogoliubov ansatz is to enable us to choose the new functions
{\bb\phhi^{_0}_{\kk\alp}\{{\bf r}\}\fb} and  {\bb\phhi^{_1}_{\kk\alp}\{{\bf r}\}\fb}
in such a way as to simplify the expression for the total effective Hamiltonian,
which will be given for a static configuration by
{\be \hHH^\prime=\int{\rm d}^3 r\hcalH^\prime\{{\bf r}\}\, ,\label{49}\fe}
with
{\be \hcalH^\prime\{{\bf r}\}= \sum_\sigma
\hpsi^{\dagh}_{\!\sigma}\!\{{\bf r}\}\calH^\prime_{_{\rm ind}}
 \hpsi_{\!\sigma}\{{\bf r}\}+
\DDelta\{{\bf r}\}\hpsi^{\dagh}_{\!\uparrow}\!\{{\bf r}\}
\hpsi^{\dagh}_{\!\downarrow}\!\{{\bf r}\}+\DDelta^\ast\!\{{\bf r}\}
\hpsi_{\!\downarrow}\!\{{\bf r}\}\hpsi_{\!\uparrow}\!\{{\bf r}\}
\, ,\label{50}\fe}
in which the independent particle contribution is given by
{\be \calH^\prime_{_{\rm ind}}= \calH_{_{\rm ind}}-\mmu
\, ,\label{51}  \fe}
where, as before, {\bb\mmu\fb} is a Lagrange multiplier, whose purpose when we apply 
the variation principle, is to impose the constraint that the expectation of the total 
integrated number density should be held fixed. It is to be remarked that in the 
presence of the pairing interaction term, the number operator {\bb \hbnn\fb} will
no longer exactly commute with the Hamiltonian, which implies that the state that 
minimises the expectation of the effective Hamiltonian obtained in this way will not 
be an exact eigenstate either of the particle number or of the energy.

The simplification of the Hamiltonian \rf{50}\fr can be achieved by taking the functions {\bb \phhi^{_0}_{\kk\alp}\{{\bf r}\}\fb}
and {\bb \phhi^{_1}_{\kk\alp}\{{\bf r}\}\fb} to be solutions of the coupled set of 
differential equations (known as the Bogoliubov-de Gennes equations in the condensed 
matter field \cite{Degennes}) given by
{\be \left({\calH^\prime_{_{\rm ind}}\atop\DDelta^\ast} {\DDelta\atop
-\calH^{\prime\ast}_{_{\rm ind}}}\right)\left(\phhi^{_0}_{\kk\alp}\atop
\phhi^{_1}_{\kk\alp}\right)=\EE_{\kk\alp}\left(\phhi^{_0}_{\kk\alp}\atop
\phhi^{_1}_{\kk\alp}\right)\, ,\label{56}\fe}
in which the eigenvalue {\bb \EE_{\kk\alp}\fb} is what will be seen to be interpretable as
the relevant quasiparticle energy. This system can be written more explicitly in 
terms of the ordinarily periodic functions {\bb\uu^{_0}_{\kk\alp}\{{\bf r}\}\fb}, 
and {\bb\uu^{_1}_{\kk\alp}\{{\bf r}\}\fb} introduced in \rf{46}\fr as
{\be \left({\calH_\kk+\VV^\prime\atop\DDelta^\ast} {\DDelta\atop-\calH^\ast_\kk
-\VV^\prime}\right)\left(\uu^{_0}_{\kk\alp}\atop\uu^{_1}_{\kk\alp}\right)=
\EE_{\kk\alp}\left(\uu^{_0}_{\kk\alp}\atop\uu^{_1}_{\kk\alp}\right)
\, ,\label{56a}\fe}
using the notation of \rf{51}\fr, where
{\be \VV^\prime= \VV -\mmu\, .\label{56b} \fe}

The foregoing specification is incomplete, because the condition of satisfying 
\rf{56}\fr will evidently be preserved by interchanges of the form
{\be \phhi^{_1\ast}_{\kk\alp}\leftrightarrow \phhi^{_0}_{-\kk\alp}\, ,
\hskip 1 cm \EE_{\kk\alp}\leftrightarrow -\EE_{\kk\alp}\, ,\label{57}\fe}
but this ambiguity is resolved by adoption of the usual postulate that
the eigenvalues be positive,
{\be  \EE_{\kk\alp}>0\, .\label{58}\fe}
To fix the normalisation of the solutions, which will automatically satisfy the 
integral relations expressible -- restoring the explicit reference to the position 
dependence -- as
{\be \int{\rm d}^3 r\,\phhi^{_1}_{\kk\alp}\{{\bf r}\}\phhi^{_0}_{\ll\bet}\{{\bf r}\}
= \int{\rm d}^3 r\,\phhi^{_0}_{\kk\alp}\{{\bf r}\}\phhi^{_1}_{\ll\bet}\{{\bf r}\}
\, ,\label{59}\fe}
the amplitude of the (automatically mutually orthogonal) solutions is chosen so that
{\be \int{\rm d}^3 r( \phhi^{_0\ast}_{\kk\alp}\{{\bf r}\}\phhi^{_0}_{\ll\bet}
\{{\bf r}\}+\phhi^{_1\ast}_{\kk\alp}\{{\bf r}\}\phhi^{_1}_{\ll\bet}\{{\bf r}\})=
\delta_{\kk\ll}\delta_{\alp\bet}\, .\label{62}\fe}

The foregoing ansatz eliminates all the undesirable terms, reducing the effective 
Hamiltonian operator to the remarkably simple form
{\be \hHH^\prime= \sum_{\sigma,\kk,\alp} \EE_{\kk\alp}\left(
\gamh^\dagh_{\!\sigma\kk\alp}\gamh_{\sigma\kk\alp}-{\rm sin}^2\theta_{\kk\alp}
\right)=\sum_{\sigma,\kk\,\alp} \EE_{\kk\alp}\left({\rm cos}^2\theta_{\kk\alp}-
\gamh_{\sigma\kk\alp}\gamh^\dagh_{\!\sigma\kk\alp}\right)\, ,\label{63}\fe}
in which {\bb \theta_{\kk\alp}\fb} is the relevant Bogoliubov angle, as defined, for 
each value of the wavenumber covector {\bb \kk_i\fb} and band index {\bb \alp\fb} by
{\be  {\rm cos}^2\theta_{\kk\alp}=\int{\rm d}^3 r\, \phhi^{_0\ast}_{\kk\alp}\{
{\bf r}\}\phhi^{_0}_{\kk\alp}\{{\bf r}\}\, ,\hskip 1 cm {\rm sin}^2\theta_{\kk\alp}=
\int{\rm d}^3 r\, \phhi^{_1\ast}_{\kk\alp}\{{\bf r}\}\phhi^{_1}_{\kk\alp}\{{\bf r}\}
\, .\label{64}\fe}

By minimisation of the expectation of the operator \rf{63}\fr one obtains the 
required condensate reference state {\bb |\,\rangle=|_{\{\mmu\}}\rangle\,\fb}, which 
is characterised by the condition
 {\be \gamh_{\sigma\kk\alp}|_{\{\mmu\}}\rangle\,=0\, ,\label{65}\fe}
expressing absence of all the quasiparticles created by the operators 
{\bb\gamh^\dagh_{\!\sigma\kk\alp}\,\fb}.

The quasiparticle operators can be written in terms of the particle operators
remembering equation \rf{21}\fr and the orthonormality condition \rf{14}\fr as
{\be  \hcc_{\uparrow\kk\alp}=\sum_{\ll,\bet} \left( U_{\kk\alp,\ll\bet}\, 
\gamh_{\uparrow\ll\bet}-\, V_{\kk\alp,\ll\bet}\gamh_\downarrow{^{\!\dagh}_{\ll\bet}} 
\right)\, , \label{65a}\fe}
{\be  \hcc_{\downarrow\kk\alp}=\sum_{\ll\bet}\left( U_{\kk\alp,\ll\bet}\, 
\gamh_{\downarrow\ll\bet} +\, V_{\kk\alp,\ll\bet}
\gamh_\uparrow{^{\!\dagh}_{_\ll\bet}}\right)\, , \label{65b}\fe}
where we have introduced the  matrices
{\be U_{\kk\alp,\ll \bet} = \int{\rm d}^3 r\, {\phhi^\ast}_{\kk\alp}\{{\bf r}\}
\phhi^{_0}_{\ll\bet}\{{\bf r}\}\, , \fe}
{\be V_{\kk\alp, \ll\bet} = \int{\rm d}^3 r\, {\phhi^\ast}_{\kk\alp}\{{\bf r}\}
\phhi^{_1\ast}_{\ll\bet}\{{\bf r}\}\, , \fe}
which from the properties of Bloch wave functions reduce to
{\be U_{\kk\alp, \ll\bet} =\delta_{\kk\ll}\, U_{\kk\alp,\kk\bet}\, , \hskip 1 cm
V_{\kk\alp, \ll\bet}=\delta_{-\kk \ll}\,V_{\kk\alp,-\kk \bet}\, .\fe}
It is to be noted that {\bb \phhi_{\kk\alp}\, ,\, \phhi^{_0}_{\kk\alp} \fb} and 
{\bb \phhi^{_1}_{\kk\alp} \fb} are all Bloch wave functions associated with the same
Bloch wave vector (hence having the same phase shift whenever translated from
one cell to another) but are solutions of different equations. Inserting these 
expressions into the number density operator
{\bb \hnn_{\sigma\kk\alp}\fb} introduced in \rf{30b}\fr, it is readily verified that 
its expectation value in the superfluid ground state is given by
{\be \langle_{\{\mmu\}}|\hnn_{\sigma\kk\alp}|_{\{\mmu\}}\rangle
=\sum_\bet |V_{\kk\alp, \,-\kk\bet}|^2\label{65d}\,  .\fe}

Remembering that {\bb \phhi_{\kk\alp} \fb} are the single particle states of the independent
Hamiltonian \rf{51}\fr, with energies {\bb \calE_{\kk\alp}^\prime\fb}, it can be
seen from the Bogoliubov equations \rf{56}\fr that the expression \rf{65d}\fr is 
equivalent to
{\be  \langle_{\{\mmu\}}|\hnn_{\sigma\kk\alp}|_{\{\mmu\}}\rangle=\sum_\bet 
\frac{|\DDelta^{_0}_{\kk\alp,  -\kk \bet}|^2}{(\EE_{-\kk\bet}+
\calE^\prime_{\kk\alp})^2}\, , \fe}
where
{\be  \DDelta^{_0}_{\kk\alp,\ll\bet}= \int{\rm d}^3 r\, \phhi^\ast_{\kk\alp}\{
{\bf r}\} \DDelta\{{\bf r}\}\phhi^{_0\ast}_{\ll\bet}\{{\bf r}\}\, .\label{65e}\fe}

\section{The BCS ansatz}

Since (particularly for the middle layers of a neutron star crust, where the 
effective mass enhancement is likely~\cite{CCH, Ch2004} to be most important) we are
still far from having a sufficient knowledge of the solutions {\bb \phhi_{\kk\alp}
\{{\bf r}\}\fb} for the independent particle model, it will evidently take some 
time before we can  hope to obtain a complete evaluation of the solutions for the 
coupled equations for {\bb \phhi^{_0}_{\kk\alp}\{{\bf r}\}\fb} and {\bb 
\phhi^{_1}_{\kk\alp}\{{\bf r}\}\fb} using an accurate estimate of the coupling 
coefficient {\bb \DDelta\{{\bf r}\}\fb}. In the meanwhile, as an immediately
available approximation, offering the best that can be hoped for as a provisional 
estimate in the short run, we can use an ansatz of the standard BCS kind, which means
adopting the prescription
{\be U_{\kk\alp, \ll\bet} ={\rm cos}\, \theta_{\kk\alp} 
\delta_{\kk\ll}\delta_{\alp\bet}  \, , \hskip 1 cm V_{\kk\alp, \ll\bet}= {\rm sin}\, 
\theta_{\kk\alp} \delta_{-\kk,\ll}\delta_{\alp\bet}\, .\label{66a}\fe}
Comparing with \rf{47}\fr and \rf{48}\fr, the Bogoliubov particle-hole doublet 
reduces to
{\be \phhi^{_0}_{\kk\alp}\{{\bf r}\}=
 \phhi_{\kk\alp}\{{\bf r}\}\,{\rm cos}\,\theta_{\kk\alp} \, ,\hskip 1 cm
\phhi^{_1}_{\kk\alp}\{{\bf r}\}=
 \phhi_{\kk\alp}\{{\bf r}\}\,{\rm sin}\,\theta_{\kk\alp} \, ,\label{66} \fe}
where the single component wave functions {\bb \phhi_{\kk\alp}\{{\bf r}\}\fb} are
the (more easily obtainable) independent particle eigenfunctions, which
can be seen from the preceding work to be specifiable as solutions of the
simple Schroedinger type equation
{\be \calH^\prime_{_{\rm ind}} \phhi_{\kk\alp}=\calE_{\kk\alp}^\prime 
\phhi_{\kk\alp}\, , \label{67}\fe}
where, in the static case under consideration at this stage, we simply have
{\be\calE_{\kk\alp}^\prime =\calE_{\kk\alp}-\mmu\label{68}\fe}
where {\bb \calE_{\kk\alp}\fb} is the ordinary Bloch energy value as introduced
in \rf{16}\fr .

It can be seen that the ansatz \rf{66}\fr will provide an exact solution
in the limit for which the relevant coupling field matrix elements
{\be \DDelta_{\kk\alp,\ll\bet}= \int{\rm d}^3 r\, \phhi^\ast_{\kk\alp}\{{\bf r}\}
\DDelta\{{\bf r}\}\phhi_{\ll\bet}\{{\bf r}\}\, ,\label{69}\fe}
reduce to diagonal form, so that we have
{\be \DDelta_{\kk\alp,\ll\bet}=\DDelta_{\kk\alp} \delta_{\kk\ll}\delta_{\alp\bet}
\, ,\label{70}\fe}
using the notation
{\be \DDelta_{\kk\alp}=\DDelta_{\kk\alp,\kk\alp}\, .\label{71}\fe}

The relation \rf{71}\fr will  be a good approximation when {\bb\DDelta_{\kk\alp}\fb} 
remains close to a fixed value {\bb\DDelta_{\rm F}\fb} (which can be taken
without loss of generality to be real and positive by choosing the
relevant phase)
in the neighbourhood of the Fermi surface, and it will evidently hold exactly when the 
coupling constant is uniform, so that  {\bb\DDelta\{{\bf r}\}= \DDelta_{\rm F}
=\DDelta_{\kk\alp}.\fb}  In the general case, as a result of the periodicity of 
{\bb \DDelta\{{\bf r}\}\fb} the pairing field matrix elements will automatically be 
diagonal in phase space, namely {\bb \DDelta_{\kk\alp,\ll\bet} = \delta_{\kk\ll}
\DDelta_{\kk\alp,\kk\bet}.\fb}  However the pairing interactions may couple single 
particle states belonging to different bands and it will only be an approximation to 
neglect those contributions when, for instance, {\bb \DDelta\{{\bf r}\}\fb} is a field 
of the radially dependent form that has been  obtained~\cite{SGL}  within the Wigner Seitz
approximation. Actually the only non vanishing matrix elements are those relating
independent single particles states belonging to the same irreducible representation
of the space group \cite{KG64}, which means that only the band states having the 
same symmetry properties may be coupled. Subject to the validity of \rf{70}\fr, the
BCS ansatz \rf{66}\fr will reduce the Bogoliubov  system of differential equations 
\rf{56}\fr to a purely algebraic eigenvalue system whose solutions have the
well-known form
{\be \EE_{\kk\alp}=\sqrt{\calE^\prime_{\kk\alp}{^2}+\DDelta_{\kk\alp}^2}\, 
,\label{73}\fe}
{\be {\rm cos}^2\theta_{\kk\alp}={\EE_{\kk\alp}+\calE^\prime_{\kk\alp}
\over2\EE_{\kk\alp}}\, ,\hskip 1 cm {\rm sin}^2\theta_{\kk\alp}={\EE_{\kk\alp}
-\calE^\prime_{\kk\alp}\over2\EE_{\kk\alp}} \, .\label{75}\fe}

It can be seen that the ansatz \rf{66}\fr has the effect of reducing 
the Bogoliubov transformation to the simple form given by
{\be  \hcc_{-\sigma\kk\alp}={\rm cos}\,\theta_{\kk\alp}\,  
\gamh_{-\sigma\kk\alp}+\sigma\,{\rm sin}\,\theta_{\kk\alp}\,
\gamh_{\sigma}{^{\!\dagh}_{-\kk\alp}}\, ,\label{76}\fe}
which is equivalent to taking
{\be \gamh_{-\sigma\kk\alp}={\rm cos}\,\theta_{\kk\alp}\,\hcc_{-\sigma\kk\alp} 
-\sigma\,{\rm sin}\,\theta_{\kk\alp}\, \hcc_{\sigma}{^{\!\dagh}_{-\kk\alp}}
\, .\label{77}\fe}

It follows from this that for the state {\bb |_{\{\mmu\}}\rangle\fb} characterised 
by \rf{65}\fr, the expectation values of the Bloch wave vector dependent number density
operators {\bb \hnn_{\sigma\kk\alp}\fb} introduced in \rf{30b}\fr will be given by
{\be \langle_{\{\mmu\}}|\hnn_{\sigma\kk\alp}|_{\{\mmu\}}\rangle={\rm sin}^2
\theta_{\kk\alp}\, ,\label{78}\fe}
This result is interpretable as expressing the effect commonly described as a
smearing of the Fermi surface, whereby the smoothed out Bloch wave vector space
distribution \rf{78}\fr replaces the hard cut off expressed by the Heaviside formula \rf{36c}\fr
that applies in limit when the pairing interaction is ignored.

\section{Formula for the mobility tensor}

When the static contribution characterised by \rf{51}\fr is extended by the 
inclusion of the current constraint term proportional to the momentum covector
{\bb \pp_i=\hba\qq_i\fb} in the effective energy \rf{33}\fr, it can be seen from
\rf{26b}\fr that as in the independent particle limit, its  effect to first order will be entirely
taken into account by merely making the gauge adjustment {\bb \aa_i = -\qq_i ,\fb} 
in the kinetic energy operator \rf{8}\fr, which means changing {\bb \kk_i \fb}  to
{\bb \kk_i - \qq_i \fb} in equation \rf{56a}\fr. The first order effect of the 
current will therefore be given, according to \rf{35a}\fr, by the adjustment
 {\be \calE^\prime_{\kk\alp}\mapsto\calE^\prime_{\{\pp\}\kk\alp}
 =\calE^\prime_{(\kk-\qq)\alp} \, ,\label{79}\fe}
for the single particle energy, and by the ensuing set of infinitesimal transformations
{\be \gamh_{\sigma\kk\alp}\mapsto \gamh_{\{\pp\}\sigma\kk\alp}\, ,\hskip 1 cm
|_{\{\mmu\}}\rangle\mapsto |_{\{\mmu,\pp\}}\rangle\, ,
\label{80}\fe}
of quasi-particle operators and state vector, while particularly, in the framework of
the BCS approximation based on the neglect of interband couplings, the Bogoliubov 
angles introduced in \rf{66}\fr will undergo a corresponding adjustment 
{\be \theta_{\kk\alp}\mapsto \theta_{\{\pp\}\kk\alp}\, .\label{80a}\fe}
As in the absence of pairing, in the strict BCS case characterised by a fixed gap 
value, {\bb\DDelta_{\kk\alp}=\DDelta_{\rm F},\fb} the result will still be describable 
just as a uniform displacement {\bb \delta\kk_i=\qq_i\fb} in the space of Bloch 
wavevectors {\bb \kk_i\fb}.

As the adjusted version  of \rf{27a}\fr, it can be seen that for any state 
{\bb |\,\rangle\fb} satisfying the simplicity condition that except for the 
diagonal contributions characterised by {\bb \sigma^\prime=\sigma, \,\ll_i 
=\kk_i \fb} and {\bb \alp=\bet\fb} the contributions of the expectation values
{\bb \langle|\hcc^\dagh_{\!\{\pp\}\sigma\kk\alp}\hcc_{\{\pp\}\sigma^\prime\ll\bet} 
|\rangle\fb} will vanish -- or  be negligible to the order of approximation under 
consideration --  the mean current defined by \rf{36h}\fr will be given for each spin 
value by the formula
{\be \bnn^i_\sigma=\sum_{\kk,\alp}  \vv_{\kk\alp}^i\,
\langle\,|\hnn_{\{\pp\}\sigma\kk\alp} |\,\rangle\, ,\label{82}\fe}
In the framework of the BCS approximation this formula will be applicable, in
particular, to the conducting reference state  {\bb |\,\rangle= |_{\{\mmu,\pp\}}
\rangle\fb}, so that by the adjusted analogue of \rf{76}\fr  the ensuing replacement of 
the formula \rf{36g}\fr, for the mean current in this state,  will be obtainable  from 
the substitution
{\be \langle_{\{\mmu,\pp\}}|\hnn_{\{\pp\}\sigma\kk\alp} |_{\{\mmu,\pp\}}\rangle=
{\rm sin}^2\theta_{ \{\pp\}\kk\alp}\, ,\label{81a}\fe}
which leads to the expression
{\be \bnn^i=\sum_\sigma \bnn^i_\sigma=2\sum_{\kk,\alp} v^i_{\kk\alp}\,{\rm sin}^2
\theta_{ \{\pp\}\kk\alp}\, .\label{81}\fe}
for the corresponding total current.

Since the total current evidently cancels out in the unperturbed static state
{\bb |_{\{\mmu\}}\rangle\, ,\fb} the quantity given by \rf{81}\fr will be expressible
to first order, in the weak current limit with which we are working, as
{\be \bnn^i=2\sum_{\kk,\alp} \vv^i_{\kk\alp}\,\pp_j{\partial ({\rm sin}^2
\theta_{ \{\pp\}\kk\alp})\over\partial \pp_j}
\, .\label{83} \fe}
The conclusion to be drawn from this is that the value of the current will be given
to linear order by an expression of the same general form
{\be \bnn^i=\pp_j\calK^{ij}\, ,\label{86}\fe}
as in the absence of pairing, but with required mobility tensor now given by an
expression of the form
{\be  \calK^{ij}=2\sum_{\kk,\alp} \vv^i_{\kk\alp}\,
{\partial ({\rm sin}^2
\theta_{ \{\pp\}\kk\alp})\over\partial \pp_j} \, .\label{87b}\fe}
It follows from \rf{79}\fr that in this small {\bb |\pp|\fb} limit we shall have
{\be {\partial \calE^\prime_{\{\pp\}\kk\alp}\over\partial\pp_i}=-
 {\partial\calE^\prime_{\{\pp\}\kk\alp}\over\hba\partial\kk_i}=-
{\partial\calE_{\kk\alp}\over\hba\partial\kk_i}=- \vv_{\kk\alp}^i\, ,\label{84}\fe}
and hence that the partial derivative in \rf{83}\fr can be evaluated
in the BCS approximation as
{\be {\partial ({\rm sin}^2\theta_{ \{\pp\}\kk\alp})\over\partial \pp_i} =
- \vv^i_{\kk\alp}{\partial ({\rm sin}^2\theta_{\kk\alp})\over\partial
\calE^\prime_{\kk\alp}} =- \vv^i_{\kk\alp}{\partial ({\rm sin}^2\theta_{\kk\alp})
\over\partial \calE_{\kk\alp}} \, ,\label{85}\fe}
in which {\bb{\rm sin}^2\theta_{\kk\alp}\fb} is given as a function of the quantity
{\bb \calE^\prime_{\kk\alp}=\calE_{\kk\alp}-\mmu\fb} and of {\bb\DDelta_{\rm F}\fb}
by \rf{75}\fr. The mobility tensor will therefore be expressible as
{\be  \calK^{ij}=-2\sum_{\kk,\alp}{\partial ({\rm sin}^2\theta_{\kk\alp})
\over \partial \calE_{\kk\alp}}\, \vv_{\kk\alp}^i \vv_{\kk\alp}^j\, ,\label{87}\fe}
in which, by \rf{75}\fr, the relevant coefficient will be given by
{\be {\partial ({\rm sin}^2\theta_{\kk\alp})
\over \partial\calE_{\kk\alp}}=-{\DDelta_{\rm F}^2\over 2 \EE_{\kk\alp}^3}
\, .\label{88}\fe}

The translation of the discrete summation formula \rf{87}\fr into the language of
continuous integration (in the limit in which the size of the mesoscopic cell is
much larger than the lattice spacing) is given by \rf{0}\fr.

Except near the base of the neutron star crust where the nuclei may acquire exotic
(e.g. ``spaghetti'' or ``lasagna'' type) configurations, it is to be expected that
the mobility tensor will have the isotropic form
{\be \calK^{ij}=\calK \gamma^{ij}\, ,\label{89}\fe}
where
{\be \calK={_1\over ^3}\gamma_{ij}\calK^{ij}=-{2\over 3}\sum_{\kk,\alp}
{\partial ({\rm sin}^2\theta_{\kk\alp})
\over \partial \calE_{\kk\alp}}v_{\kk\alp}^2\, ,\hskip 1 cm \vv_{\kk\alp}^2
=\gamma_{ij} \vv_{\kk\alp}^i \vv_{\kk\alp}^j\, ,\label{90}\fe}

It is to be observed that subject to the BCS approximation of uniform coupling, meaning
that there is a constant gap parameter, {\bb\DDelta_{\kk\alp}=\DDelta_{\rm F}\fb},
the formulae \rf{87}\fr and \rf{90}\fr will be convertible, using integration by parts,
to the form
{\be  \calK^{ij}=2\sum_{\kk,\alp}{ {\rm sin}^2\theta_{\kk\alp}\over\hba^2}
{\partial^2\calE_{\kk\alp}\over \partial \kk_i \partial \kk_j}\, ,\hskip 1 cm
 \calK={_2\over^3}\sum_{\kk,\alp}{ {\rm sin}^2\theta_{\kk\alp}\over\hba^2}
\,\gamma_{ij}{\partial^2\calE_{\kk\alp}\over\partial\kk_i\partial\kk_j}
\, .\label{92}\fe}

This latter formula is useful for the evaluation of the corresponding effective
mass {\bb\mm_\star\fb} as defined  by
{\be \mm_\star=\nn/\calK\, ,\label{93}\fe}
in terms of the relevant total particle number density as given by the prescription
{\be \nn=\sum_\sigma\langle|\hbnn_\sigma|\rangle=2\sum_{\kk,\alp}
{\rm sin}^2\theta_{\kk\alp}\, ,\label{94}\fe}
in which, if we only wish to count unbound neutrons, the summation should be taken
only for values above a lower cut off below which the states are bound so that the
corresponding values of the velocity {\bb v_{\kk\alp}\fb} will vanish.

The concept of an effective mass has traditionally been a source of confusion as
different definitions have been used in different contexts. Moreover in solid state
physics one is often more interested in electric charge (not mass) whose transport is
related to the electric field {\bb \Elec_i \fb} via an Ohm type law as
{\be \jj^i = \ee\, \nn^i = {\sigm}^{ij} \Elec_j, \fe} where {\bb \nn^i \fb} is the
electron current density, and  {\bb {\sigm}^{ij} \fb} is the relevant electric
conductivity tensor. While this conductivity tensor {\bb {\sigm}^{ij} \fb} has the
advantage of relating macroscopic measurable quantities, it depends on the dynamical
evolution of the medium unlike the newly introduced mobility tensor
{\bb \calK^{ij} \fb}, on which the effective mass {\bb \mm_\star \fb} is defined.
The electric conductivity tensor will be given by an expression of the form
{\be {\sigm}^{ij}=\ee^2 \tau \calK^{ij}\fe}
in which {\bb \tau \fb} is a timescale characterising the rate of decay (by various
scattering processes) towards the zero current state that is the only locally stable
configuration in the ``normal'' case. The kind of superconducting case with which we are
concerned may be described as a limit in which the relevant timescale {\bb \tau \fb} is
infinite, so that the conductivity {\bb {\sigm}^{ij}\fb} will also be infinite, even
though the mobility tensor {\bb \calK^{ij} \fb} has a well behaved finite value as in
the ``normal'' case. However it is important to understand that the reason why the
relevant timescale {\bb \tau \fb} is effectively infinite in the superconducting case
is {\it not} because relevant scattering cross sections are small (as in the case of
a ``normal'' good conductor) but rather because the current carrying
configuration is locally {\it stable} (with respect to scattering processes that may
be quite strong) in a superconducting state, for reasons that will be reviewed
in the next section.

It is to be remarked that the formula for the mobility tensor \rf{87}\fr is very
similar to the formula obtained without pairing correlations, the Heaviside unit step
distribution being merely smeared. In particular the same velocities appear
in these formulae. One might have naively guessed that apart from the particle
state distribution which is smoothed, the relevant velocity would have been given not
by the ordinary group velocity {\bb \vv^i_{\kk\alp} \fb} given as the momentum space
gradient of the energy distribution {\bb\calE_{\kk\alp}\fb} by \rf{32}\fr but by the
analogously defined quantity {\bb \tilde{\vv}^i_{\kk\alp}\fb} obtained by
substituting {\bb \EE_{\kk\alp}\fb} in place of {\bb \calE_{\kk\alp}\, ,\fb} namely
{\be \tilde \vv^i_{\kk\alp}=\frac{1}{\hba}{\partial\EE_{\kk\alp}\over\partial \kk_i}
\, .\label{96}\fe}
Actually this latter ``pseudovelocity'' is interpretable as a mean velocity between
particles and holes, since {\bb \EE_{\kk\alp} \fb} is the energy of a quasiparticle
which is a mixture of particles and holes. More specifically, when (as in the simple
BCS case for an homogeneous system) the gap parameter is independent of the momentum, this modified velocity
will be given by the expression
{\be \tilde{\vv}^i_{\kk\alp} = \vv^i_{\kk\alp}\frac{\calE^\prime_{\kk\alp}}
{\EE_{\kk\alp}} \, ,\label{97}\fe}
from which it can be seen that {\bb \tilde\vv^i_{\kk\alp} \fb} will vanish  at the
Fermi surface characterised by {\bb \calE_{\kk\alp}=\mmu \fb}, where the number of
particles is equal to the number of holes.

In the limit for which, in so far as the unbound neutrons are
concerned, the effect of the crustal nuclei is small (either because the nuclei
occupy only a small part
of the volume, as will be the case just above the neutron drip transition, or
because the nuclear surface is very diffuse, as will be the case near the base of the crust) we
shall have
{\be \frac{1}{\hba^2}{\partial^2\calE_{\kk\alp}\over\partial\kk_i
\partial\kk_j}={1\over \mm^{_\oplus}}\gamma_{ij}\, ,\label{95}\fe}
where {\bb \mm^{_\oplus}\fb} is the uniform mass scale appearing in the
kinetic energy operator, which will be comparable with, but for precision somewhat
less that, the ordinary neutron mass {\bb \mm\fb}. It can be seen by
comparing \rf{92}\fr and \rf{94}\fr that in this approximately uniform limit
the effective mass for the unbound neutrons will be given simply by
{\bb \mm_\star=\mm^{_\oplus}\fb}, regardless of whatever the value of the gap
parameter {\bb\DDelta\fb} may be.

\section{Superconductivity property and critical current}

An unsatisfactory feature of the rather profuse contemporary literature dealing
with various kinds of what is commonly referred to as ``superconductivity'' in
astrophysically relevant contexts (including such exotic varieties as colour
superconductivity in quark condensates) is the rarity of any serious
theoretical consideration of the actual property of superconductivity in the
technical sense, meaning the possibility of having a relatively moving current that
is effectively stable, or in stricter terminology {\it metastable}, with respect to
small perturbations -- such as would normally give rise to a dissipative damping
mechanism of a resistive or viscous kind.

In the astrophysical literature concerned with pulsars it has generally been taken
for granted that neutron currents of the kind considered in the present work
actually are characterised by superconductivity in the sense of being metastable
with respect to relevant kinds of perturbation. In this section we shall investigate
the conditions under which this supposition of metastability is indeed valid.
The issue is that of the stability, for small but finite values of the
momentum covector {\bb\pp_i\fb}, of the superconducting reference state
{\bb \vert\,\rangle= \vert_{\{\mmu,\pp\}}\rangle \fb} that is characterised by the
minimisation of the combination \rf{33}\fr.

The conducting state {\bb \vert_{\{\mmu,\pp\}}\rangle\fb} was derived  by minimising
the energy expectation  {\bb\langle|\hHH|\rangle\fb} subject to the condition that
the particle number expectation {\bb\langle |\hbnn|\rangle\fb} and the current
expectation {\bb\langle |\hbnn^{\, i}|\rangle\fb} were held fixed. It is physically
reasonable to suppose the particle number expectation {\bb\langle |\hbnn|\rangle\fb}
really will be preserved under the conditions of chemical equilibrium that are
envisaged in the relevant applications, but no such consideration applies to the
current expectation {\bb\langle |\hbnn^{\, i}|\rangle\fb} which in a ``normal'' state
would tend to be damped down by many conceivable kinds of scattering process. The
physically pertinent question is therefore whether {\bb \vert_{\{\mmu, \pp\}}
\rangle\fb} will still minimise {\bb\langle|\hHH|\rangle\fb} with respect to small
relevant perturbations -- subject of course to the preservation of the particle
number expectation {\bb\langle |\hbnn|\rangle\fb} as before -- when the prior
assumption of preservation of {\bb\langle |\hbnn^{\, i}|\rangle\fb} is abandonned.
Subject to the particle number conservation condition
{\be \delta\langle |\hbnn|\rangle=0\, ,\label{98}\fe}
this stability requirement is equivalent to the condition
of minimisation of  {\bb\langle|\hHH^\prime|\rangle\fb}
meaning that any admissible perturbation must satisfy
{\be \delta\langle|\hHH^\prime|\rangle >0\, ,\label{99}\fe}
where according to the notation introduced in \rf{33b}\fr,
{\be \hHH^\prime= \hHH_{\{\pp\}}^\prime+\pp_i\, \hbnn^{\, i}\, .\label{100}\fe}

According to the reasonning of the previous section, the relevant adjustment of
\rf{64}\fr will give us
{\be \hHH^\prime_{\{\pp\}}= \sum_{\sigma,\kk,\alp} \EE_{\{\pp\}\kk\alp}\left(
\gamh^\dagh_{\!\{\pp\}\sigma\kk\alp}\gamh_{\{\pp\}\sigma\kk\alp}
-{\rm sin}^2\theta_{\{\pp\}\kk\alp}\right)\, ,\label{101}\fe}
so that the specifications \rf{100}\fr and \rf{82}\fr provide the variation formula
{\be \delta\langle|\hHH^\prime|\rangle= \sum_{\sigma,\kk,\alp}
\left( \EE_{\{\pp\}\kk\alp}\delta\langle |\gamh^\dagh_{\!\{\pp\}\sigma\kk\alp}
\gamh_{\{\pp\}\sigma\kk\alp}|\rangle+\pp_i\,\vv^i_{\kk\alp}\delta
\langle|\hnn_{\{\pp\}\sigma\kk\alp}|\rangle\right)\, ,\label{102}\fe}
in which, for the BCS case, it can be seen from \rf{73}\fr that we shall have
{\be \EE_{\{\pp\}\kk\alp}=\sqrt{\calE^\prime_{\{\pp\}\kk\alp}{^2}+\DDelta_{\kk\alp}^2}
\, .\label{103}\fe}
In this BCS case, the action on the conducting state {\bb |_{\{\mmu,\pp\}}\rangle\fb}
of a typical quasiparticle
creation operator {\bb \gamh^\dagh_{\!\{\pp\}\uparrow\kk\alp}\fb} will provide
only three non vanishing terms in the sum \rf{102}\fr, namely those given by
{\be \delta\langle |\gamh^\dagh_{\!\{\pp\}\uparrow\kk\alp}
\gamh_{\{\pp\}\uparrow\kk\alp}|\rangle=1 \, ,\label{104}\fe}
together with the number variation contributions
{\be\delta \langle|\hnn_{\{\pp\}\uparrow\kk\alp}|\rangle={\rm cos}^2
\theta_{\{\pp\}\kk\alp} \, ,\hskip 1 cm \delta \langle|\hnn_{\{\pp\}\downarrow-\kk\alp}
|\rangle =- {\rm sin}^2\theta_{\{\pp\}-\kk\alp}\, .\label{105}\fe}
It follows from the symmetry properties {\bb \vv^i_{\kk\alp}
=-\vv^i_{-\kk\alp}\fb} and {\bb \theta_{\kk\alp}=\theta_{-\kk\alp}\fb}
that the explicit dependence on the Bogoliubov angle will cancel out at first order
in the net energy contribution provided in \rf{102}\fr by such an excitation, so that
this energy contribution will be positive if and only if
{\be \EE_{\{\pp\}\kk\alp} +\pp_i\,\vv^i_{\kk\alp}  >0\, .\label{106}\fe}
It is to be noted that such an individual quasiparticle excitation will in general
violate the requirement \rf{98}\fr to the effect that the number of real particles
should be conserved, but it is evident from \rf{105}\fr that such violations
may have either sign and so can be cancelled out by the combined effect of two or
more elementary excitations. What, in a stable case, can not be cancelled out is the
combined effect of several quasiparticle energy contributions in \rf{102}\fr: it can
be seen  that the quasiparticle energy contributions will always add up to give the
positive result needed for stability provided the inequality \rf{106}\fr is
satisfied for all admissible modes.

The stability condition \rf{106}\fr that we have derived for a BCS type inhomogeneous
superconductor is consistent with Landau's classical treatment based on vaguer
heuristic arguments in the context of superfluid Helium 4 \cite{LL58}.
Since
the quasiparticle energy {\bb \EE_{\{\pp\}\kk\alp}  \fb} is always positive, it is
clear that the stability condition \rf{106}\fr will always be satisfied if
{\bb \pp_i\,\vv^i_{\kk\alp}  >0. \fb}  Therefore an instability in the superfluid
conducting state can only occur when {\bb \pp_i\,\vv^i_{\kk\alp}  <0. \fb} Rewriting the
inequality \rf{106}\fr as {\bb  \EE_{\{\pp\}\kk\alp} > - \pp_i\,\vv^i_{\kk\alp}, \fb}
squaring both sides and substituting the expression \rf{35}\fr
for {\bb \calE^\prime_{\{\pp\}\kk\alp}\fb} in \rf{103}\fr we see that in the BCS
case there is a remarkable simplification (which does not seem to have been pointed
out before) whereby the terms that are non linearly dependent on  the momentum
covector {\bb \pp_i\fb} cancel out,  so that the superfluid conducting state will be
stable if the following inequality is satisfied
{\be 2\,\pp_i\, \vv_{\kk\alp}^i\,\calE^\prime_{\kk\alp}<\EE^{\, 2}_{\kk\alp}
\, .\label{107}\fe}
This can be rewritten in terms of the ``pseudovelocity'' introduced in \rf{96}\fr as
{\be \pp_i\, \tilde \vv_{\kk\alp}^i<{_1\over^2}\, \EE_{\kk\alp}\, ,\label{108}\fe}
which simplifies to the following form whenever the BCS gap is independent of
the momentum
{\be \pp_i\,{\partial\over\partial \kk_i}\left({\rm ln} \{ \EE^{\, 2}_{\kk\alp} \}
\right) < \hba\, .\label{109}\fe}

The inequality \rf{107}\fr is evidently verified for elementary excitations above the
Fermi level for which {\bb \calE^\prime_{\kk\alp} >0 .\fb} In the derivation of the
inequality \rf{107}\fr we have assumed that {\bb \pp_i \vv_{\kk\alp}^i <0. \fb} Actually
there will always be some modes for which this is satisfied since whenever {\bb \pp_i
\vv_{\kk\alp}^i >0 \fb} we shall have {\bb \pp_i \vv_{-\kk\alp}^i <0 \fb}.

The stability condition \rf{106}\fr (which has to be satisfied for all modes)
can therefore be restated as the requirement
that the magnitude {\bb \pp\fb} of the mean particle momentum covector
{\bb \pp_i\fb} lies below some critical threshold {\bb \pp_{\rm c} \fb}
{\be \pp<\pp_{\rm c}\label{110}\fe}
in which, for an approximately isotropic distribution depending only on the
magnitude {\bb\kk\fb} of the wavenumber covector {\bb\kk_i\fb}, the critical value
{\bb \pp_{\rm c}\fb} will be given by
{\be \pp_{\rm c}\approx {{\rm min}\atop ^{\kk,\alp}} \left\{ {1\over 2\vv_{\kk\alp}}
\left(|\calE^\prime_{\kk\alp}|+{\DDelta_{\kk\alp}^2\over|\calE^\prime_{\kk\alp}|}
\right)\right\}\, .\label{111}\fe}

Since {\bb \calE^\prime_{\kk\alp}\fb} vanishes on the Fermi surface, it is clear
from \rf{111}\fr that  {\bb \pp_{\rm c} \fb} will also vanish -- so that there will be
no phenomenon of superconductivity -- not only when the gap {\bb\DDelta_{\kk\alp}\fb}
vanishes everywhere, but even when it vanishes just in the neighbourhood of the Fermi
surface. When the mean value {\bb \DDelta_{\rm F}\fb} of the gap at the Fermi surface is non-zero
but small compared with the other relevant energy scales -- as will typically be the
case -- it can be seen that the minimum in \rf{111}\fr will be attained for energy
values differing from the Fermi value by a small but finite positive or negative amount
that will be given approximately by {\bb\calE^\prime_{\kk\alp}\approx \pm
\DDelta_{\rm F}\fb}. In such a case, it follows that the critical momentum value
\rf{111}\fr will be expressible in terms of the mean value {\bb \vv_{\rm F}\fb} of the group
velocity magnitude {\bb \vv_{\kk\alp} \fb} at the Fermi surface by the approximation
{\be \pp_{\rm c}\approx {\DDelta_{\rm F}\over \vv_{\rm F}   }\, .\label{113}\fe}

Introducing the critical velocity as {\bb \vv_{\rm c} \equiv \pp_{\rm c}/\mm_\star \, ,\fb}
the criterion \rf{113}\fr can be written as
{\be \vv_{\rm c}\approx {\DDelta_{\rm F}\over \mm_\star \vv_{\rm F} }\, .\label{113a}\fe}
In the limit of an homogeneous
medium for which {\bb \mm_\star = \mm^{_{\oplus}}\, , \fb}
the critical velocity reduces to an expression which is commonly found in the literature concerning
homogeneous electron superconductivity in metals \cite{Bardeen62}, namely
{\be \vv_{\rm c}^{_{(0)}}\approx \frac{\DDelta_{\rm F}^{^{(0)}}}{\hba \kk_{\rm F}}\, ,\label{113b}\fe}
(using the superscript {\bb ^{_{(0)}} \fb} to indicate what would be obtained for uniform values of the
microcoscopic effective mass {\bb \mm^{_\oplus} \fb} and potential {\bb \VV \fb} )
where {\bb \kk_{\rm F} \fb} is the radius of the Fermi sphere.
It must be emphasized however that
the critical velocity of an inhomogeneous superfluid (such as the neutron superfluid in the inner layers of a neutron
star crust) will differ from the estimate \rf{113b}\fr by a factor
{\be \frac{\vv_{\rm c}}{\vv_{\rm c}^{_{(0)}} }= \frac{\DDelta_{\rm F}}{\DDelta_{\rm F}^{^{(0)}}}
\frac{S_{\rm F}}{S_{\rm F}^{^{(0)}}}\, ,\label{113c}\fe}
where {\bb S_{\rm F} \fb} and {\bb S_{\rm F}^{^{(0)}} \fb} are the Fermi
surface areas in the inhomogeneous and homogeneous cases respectively.
Since the opening of band gaps in the single particle energy spectrum
decreases the Fermi surface area, the critical velocity is therefore expected
to be smaller than the expression \rf{113b}\fr
assuming that {\bb  \DDelta_{\rm F}\approx \DDelta_{\rm F}^{^{(0)}} \, . \fb}

For a gap of the order of an MeV, in a region where the kinetic contribution to
the Fermi energy has a typical value of the order of a few tens of MeV, the
formula \rf{113}\fr implies a critical momentum value corresponding to a kinetic
energy of relative motion of the order of hundreds of keV per neutron. This is
comparable with the total kinetic energy of rotation in the most rapidly rotating
pulsars. However the relative rotation speeds of the neutron currents that are
believed to be involved in pulsar glitch phenomena are very much smaller -- by
factors of 10$^{-3}$ or even far less \cite{Ru92} -- than the absolute rotation speeds of the
neutron star. In all such cases it may therefore be concluded that the superfluidity
criterion \rf{113}\fr will be satisfied within an enormous confidence margin.

It is to be remarked that in the more thoroughly investigated context of laboratory
superfluidity ~\cite{LL58} Landau's simple linear formulation of the stability
problem in terms just of phonons provides only an upper limit on the critical
momentum whose true value is considerably reduced by the less mathematically
tractable -- since essentially non linear -- effect of rotons. Analogous
considerations presumably apply in the present context. This means that although our
present treatment places the estimate \rf{113}\fr on a sounder footing than was
provided by any previous work of which we are aware, it should still be considered
just as an upper bound on the true critical value which is likely to be substantially
reduced by non linear effects whose mathematical treatment is beyond the scope of
the methods used here. Despite this caveat, the prediction of genuine
superconductivity in the context of glitching neutron star crusts should be
considered to be very robust. The justification for such confidence is
that -- according to the considerations outlined in the preceding paragraph --
the relevant magnitudes of the neutron currents in question correspond to
values of the neutron momentum {\bb \pp\fb} that are extremely small compared with
the order of magnitude given by \rf{113}\fr. For such very low amplitude currents
there is no obvious  reason to doubt the validity of conclusions -- including
estimates of effective masses, as well as the prediction of genuine superfluidity --
that are based on the simple kind of linearised treatment used here.

\section{Conclusions}

In the middle layers of the crust, where the effect of inhomogeneities
will be important, our previous analysis neglecting the
effect of the superfluid gap lead to the prediction~\cite{CCH} that there will be a
strong ``entrainment'' effect whereby the value of {\bb \mm_\star\fb} will become
very large compared with {\bb \mm\fb}. This prediction has now been confirmed by an
analysis \cite{Ch2004} based on the phenomenological model of Oyamatsu {\it et al.}
\cite{OyYam94} where values as large as {\bb \mm_\star/\mm \sim 15\fb} has been found
at a baryon density {\bb \nn_b=0.03\, {\rm fm}^{-3} .\fb} Our present analysis
indicates that this conclusion will not be significantly affected by taking the
relevant pairing gap {\bb\DDelta\fb} into account. Thus although the pairing is essential
for the actual phenomenon of superconductivity, on the  other hand, in so far as the
effective mass is concerned, neglect of the effect of superconductivity pairing will indeed be
justifiable as a robust first approximation, at least for moderate values of the gap
parameter compared to the kinetic contribution to the Fermi energy.

The unimportance of pairing from the point of view of entrainment, which has been
usually assumed (see for instance Borumand \textit{et al.} \cite{BJK96}) and explicitely
shown in the present work, can be seen from the consideration that, when {\bb\DDelta_{\alp\kk}\fb}
is small compared to the Fermi energy {\bb \mmu\, ,\fb} the coefficient \rf{88}\fr will be very small
everywhere except in a thin layer with width of the order of {\bb\DDelta_{\alp\kk}\fb}
near the Fermi surface locus where {\bb \calE_{\alp\kk}=\mmu,\fb} which means that
when the coupling is weak its effect will be entirely negligible. In sensitive cases
for which the geometry of the energy contours near the Fermi surface is complicated
by band effects, a moderately strong pairing effect might make a significant
difference by smoothing out variations of the mobility tensor as a function of
density, but does not seem likely that this smearing effect would make much
difference to the large scale average properties of the mobility tensor. In other words
the effective mass is expected to be much more sensitive to band gaps than to the
pairing gap. The main reason is that {\bb \DDelta \fb} appears only in the number
density distribution whereas band gaps (resulting from Bragg scattering of dripped
neutrons by crustal nuclei) have a strong influence upon the neutron velocity {\bb \vv^i_{\alp\kk}
\fb} which is vanishingly small in this case.

 \bigskip
{\bf Acknowledgements}
\medskip

The authors wish to thank D.G. Yakovlev, M. Gusakov, N. Van Giai and N. Sandulescu for instructive 
conversations. The present work has been carried out within the framework of the 
European Associated Laboratory ``Astrophysics Poland-France''. Nicolas Chamel acknowledges support
from the Lavoisier program of the French Ministry of Foreign Affairs.

\vfill\eject

\end{document}